\documentclass[bibyear]{aa}

\usepackage{amsmath,amssymb}
\usepackage{mathtools}
\usepackage[utf8]{inputenc}
\usepackage[T1]{fontenc}
\usepackage{graphicx}
\usepackage{booktabs}
\usepackage{xcolor}
\usepackage{natbib}
\usepackage{hyperref}
\hypersetup{
  colorlinks = true, 
  urlcolor  = blue, 
  linkcolor = red,  
  citecolor = blue  
}

\usepackage{placeins}
\usepackage[varg]{txfonts}
\usepackage{orcidlink}
\usepackage{etoolbox}

\makeatletter
\newcommand*\aa@thmlabelfix{\protected@xdef\@currentlabel{\@currentlabel}}
\expandafter\patchcmd\csname\string\theorem\endcsname
  {\refstepcounter{\csname aa@theorem@\@currenvir @counter\endcsname}}
  {\refstepcounter{\csname aa@theorem@\@currenvir @counter\endcsname}\aa@thmlabelfix}
  {}{\GenericError{}{aa theorem label patch failed}{}{}}
\expandafter\patchcmd\csname\string\proposition\endcsname
  {\refstepcounter{\csname aa@theorem@\@currenvir @counter\endcsname}}
  {\refstepcounter{\csname aa@theorem@\@currenvir @counter\endcsname}\aa@thmlabelfix}
  {}{\GenericError{}{aa proposition label patch failed}{}{}}
\makeatother

\title{Information Geometry meets Functional ANOVA:
An Exact Fisher-Information Decomposition, with an Application to Radio Luminosity Functions}
\titlerunning{Information Geometry meets Functional ANOVA}
\author{Marko Imbri\v{s}ak\thanks{\emph{marko.imbrisak@gmail.com}}\inst{1}\orcidlink{0000-0002-2773-8617}
\and Kre\v{s}imir Tisani\'c\inst{1}\orcidlink{0000-0001-6382-4937}}

\institute{Independent researcher, Zagreb, Croatia}
\date{\today}
\abstract{
Information geometry represents a fitted model as a manifold whose metric is the Fisher information.
We show that, for a nonlinear regression model, this metric decomposes exactly across the
Hoeffding--Sobol' (functional ANOVA) channels of its covariates. Treating the score contributions of
main effects and interactions as vectors in a Hilbert space, their Gram matrix reproduces the
(Gauss--Newton) Fisher information metric exactly, and this identity extends to an exact, additive
decomposition of the Fisher information matrix into main-effect, interaction, and cross-channel
terms: a Sobol'-type accounting for \emph{parameter information} rather than output variance, and
equivalently a channel-resolved pullback of the ambient Fisher metric along the model embedding. We
give the exact decomposition theorem with a short proof (deriving both the centered and uncentered
Fisher information), define information-theoretic analogues of Sobol' indices, characterize the sign
and finite-sample behaviour of the cross-channel terms, and describe two practical estimators. As an
astrophysical application we decompose the parameter information of a combined 1.4~GHz radio
luminosity-function model spanning active galactic nuclei (AGN) and star-forming galaxies (SFG), fit
to 6dFGS--NVSS and VLA-COSMOS 3~GHz data: the decomposition localizes where in luminosity each
population and survey constrains the fit, and exposes the compensating (``sloppy'') channel
combinations near the luminosity-function turnover. A plant CO$_2$-uptake experiment serves as a
small, fully categorical companion example.}


\begin{document}
\maketitle
% linenoaa.sty appends \linenumbers to \maketitle, so switch them off here
% (after \maketitle) for the arXiv version.
\nolinenumbers

\section{Introduction}
\label{sec:intro}
Functional ANOVA (Hoeffding--Sobol') decomposition of a model's output is a standard tool for global
sensitivity analysis: a function of independent inputs is written as a sum of a constant, main
effects, and interactions, orthogonal in $L^2$ under input independence \citep{hoeffding1948,sobol1993}. Extensions to
dependent inputs exist \citep{stone1994,hooker2007,chastaing2012}, including a recent closed-form construction for general marginals
\citep{ferrere2026}.

This literature decomposes the \emph{output variance} of a fixed function. We ask a related but
distinct question, motivated by uncertainty quantification for nonlinear model fitting with
heterogeneous, possibly dependent covariates: how does the \emph{Fisher information} about the
model's parameters decompose across the same ANOVA channels? Variance and information are different
functionals of a model, and, to our knowledge, the exact decomposition we derive
(Theorem~\ref{thm:fs}) has not been stated in this form in either the sensitivity-analysis or the
identifiability literature. Our contribution is narrow and specific: an exact identity linking the
Gram matrix of Hoeffding--Sobol' score components to the Fisher information metric; the general
decomposition theorem, with information-theoretic Sobol' indices and an account of the sign and
finite-sample behaviour of the cross terms; and two illustrations on categorical-covariate datasets.

The relation to existing work is as follows. The Hoeffding--Sobol' machinery of
Section~\ref{sec:background} is classical \citep{hoeffding1948,sobol1993}; its extension to dependent inputs via hierarchical
orthogonality is due to \citet{stone1994,hooker2007}, formalized by \citet{chastaing2012}, and given a closed-form basis very recently by \citet{ferrere2026}.
None of this prior work decomposes the Fisher information of a fitted model. The closest existing
tools are derivative-based global sensitivity measures \citep[DGSM;][]{sobol2009} and active-subspace
methods \citep{constantine2015}, which aggregate output gradients over the input space to rank inputs
or find dominant input directions, but both decompose properties of the \emph{output} (its variance
or its gradient energy), not the Fisher information of the \emph{parameters}, and neither carries the
exact channel-additivity of Theorem~\ref{thm:fs}. That specific construction is, as far as we have
been able to determine, new. The geometric
reading of the Fisher information metric as an embedding metric, and its second derivative as
extrinsic/parameter-effects curvature, follows \citet{bates1980}; the negative cross-channel terms are the
information-geometric signature of the sloppy parameter combinations of \citet{transtrum2010,transtrum2011}. At root the
construction rests on the same fact from the theory of curved exponential families \citep{efron1975}: the Fisher
information for a parameter path embedded in a larger exponential family is the pullback of the
ambient Fisher metric \citep{barndorff1978} along that embedding \citep{amari1985,kass1997}; our contribution is to index this pullback
by Hoeffding--Sobol' ANOVA channels rather than by generic parameter blocks. We deliberately do not
claim novelty for the underlying ANOVA machinery itself.

\section{The Fisher-Sobol decomposition}
\label{sec:decomp}

\subsection{Hoeffding--Sobol' components as Hilbert-space geometry}
\label{sec:background}
Let $Y\in L^2$ and let $X_1,X_2$ be covariates. Writing $M_i=L^2(\sigma(X_i))$, the conditional
expectation $\mathbb E[Y\mid X_i]$ is the orthogonal projection of $Y$ onto $M_i$. After centering,
$g_i(X_i):=\mathbb E[Y\mid X_i]-\mathbb E[Y]$ is a \emph{vector} in $L^2$, not a scalar; for two
covariates, with $g_{12}:=\mathbb E[Y\mid X_1,X_2]-\mathbb E[Y]-g_1-g_2$, one has under independence
the orthogonal (Hoeffding) decomposition
\begin{equation}
\begin{aligned}
Y &= \mathbb E[Y] + g_1(X_1) + g_2(X_2) + g_{12}(X_1,X_2) + \varepsilon,\\
\|\cdot\|^2 &\text{ additive}.
\end{aligned}
\label{eq:hoeffding}
\end{equation}
The object $g_{12}$ is a vector, not a metric; the Gram matrix $G_{uv}=\langle g_u,g_v\rangle$,
$u,v\in\{1,2,12\}$, is symmetric positive semi-definite and defines a metric on the space of ANOVA
components of a chosen model class, not on the raw covariate space, which remains two-dimensional
even though $G$ is $3\times3$. When $X_1,X_2$ are dependent, $G$ acquires off-diagonal structure that
must be removed by explicit Gram--Schmidt orthogonalization to recover a hierarchically orthogonal
(in the sense of \citep{hooker2007}) set of components.

\subsection{Exact link to the Fisher information metric}
\label{sec:identity}
Consider a regression model $f(X_1,X_2;\theta)$ fit to data $(X_{1k},X_{2k},Y_k)_{k=1}^N$ by weighted
nonlinear least squares, with studentized residual $r_k(\theta)=[Y_k-f_k(\theta)]/\sigma_k$, where
$\sigma_k$ is the per-point noise level. For Gaussian noise the Fisher information metric is the
Gauss--Newton matrix $g_{\mu\nu}=\sum_k\partial_\mu f_k\,\partial_\nu f_k/\sigma_k^2$. In the
homoscedastic case $\sigma_k\equiv\sigma$, if $f$ is written in the centered, hierarchically orthogonal
basis $f=\mu+\theta_1\phi_1(X_1)+\theta_2\phi_2(X_2)+\theta_{12}\phi_{12}(X_1,X_2)$, then
\begin{equation}
\begin{aligned}
g_{\theta^a\theta^b} &= \frac{N}{\sigma^2}\,\hat G^\phi_{ab},\\
\hat G^\phi_{ab} &:= \frac1N\sum_k \phi_a^{(k)}\phi_b^{(k)},
\end{aligned}
\label{eq:main-identity}
\end{equation}
i.e.\ the parameter-space Fisher information matrix is exactly $N/\sigma^2$ times the (bare) Gram
matrix of the basis functions. We adopt the conventional unit-noise normalization $\sigma^2=1$
throughout; for a general \emph{homoscedastic} noise level the explicit factor $1/\sigma^2$ is carried
as above, while for \emph{heteroscedastic} noise $\varepsilon_k\sim\mathcal N(0,\sigma_k^2)$ the metric
becomes the weighted Gram matrix $g_{\theta^a\theta^b}=\sum_k w_k\,\phi_a^{(k)}\phi_b^{(k)}$ with
$w_k=1/\sigma_k^2$ (equivalently, a diagonal weight $W=\operatorname{diag}(w_k)$ in the pullback
$\partial f^\top W\,\partial f$). Since every quantity of interest below is a ratio of Fisher-Sobol
channels, the overall noise scale cancels and the normalization is immaterial for the decomposition;
we state it only to fix conventions. Because the $\phi_a$ are centered, the intercept $\mu$ decouples
exactly from $(\theta_1,\theta_2,\theta_{12})$ in this basis (the linear reparametrization from raw
monomial coefficients has Jacobian determinant~1, so no information is created or destroyed by the
change of basis).

Identity~\eqref{eq:main-identity} is an instance of a general fact about curved exponential families:
for $Y_k=f_k(\theta)+\varepsilon_k$, $\varepsilon_k\sim\mathcal N(0,\sigma^2)$, itself the canonical
example of a curved exponential family \citep{efron1975}, the Fisher information for $\theta$ is the pullback of
the ambient Fisher metric along the embedding $\theta\mapsto f(\theta)$, $g_{\mu\nu}=\partial_\mu
f^\top\Sigma^{-1}\partial_\nu f$ with $\Sigma=\sigma^2 I$. In the linear-in-$\theta$, hierarchically
orthogonal basis of Section~\ref{sec:background}, $\partial_a f=\phi_a$, and the pullback reduces
exactly to $(N/\sigma^2)\hat G^\phi_{ab}$, connecting the construction to the classical geometry of
nonlinear regression \citep{bates1980} and to information geometry more broadly \citep{barndorff1978,amari1985,kass1997}. A second, distinct
object, the attribution Gram matrix $\hat G_{ab}=\theta_a\theta_b\hat G^\phi_{ab}$ (coefficients
included), answers a different question (how much of $\mathrm{Var}(\hat Y)$ is attributable to channel
$a$) and must not be conflated with $g_{\theta\theta}$; the two coincide only up to the diagonal
rescaling by $\theta_a\theta_b$.

\subsection{Exact decomposition theorem}
\label{sec:theorem}
For a general nonlinear $f(X_1,X_2;\theta)$, let $h_u(X;\theta)$, $u\in\{0,1,2,12\}$ (mean, main
effects, interaction) be the Hoeffding--Sobol' components of $f$, estimated either nonparametrically
or by projection onto an orthogonal basis (Section~\ref{sec:estimators}). Provided the covariate
distribution is held fixed (independent of $\theta$) and the model satisfies the mild regularity
permitting interchange of $\partial_\theta$ with the conditional-expectation / orthogonal-projection
operators defining the $h_u$ (dominated convergence, e.g.\ $\partial_\theta f$ and $f$ jointly
integrable in a neighbourhood of $\theta$), differentiation and decomposition commute exactly,
$\partial_\theta[h_u(f)]=h_u(\partial_\theta f)$, since the $h_u$ are fixed bounded linear projection
operators on $L^2(\sigma(X))$ and linear operators commute with $\partial_\theta$. Define
\begin{equation}
T^{uv}_{ab} := \sum_k \partial_a h_u^k\,\partial_b h_v^k.
\label{eq:Tdef}
\end{equation}

\begin{theorem}[Exact Fisher-Sobol decomposition]
\label{thm:fs}
Let $g_{ab}=\sum_k\partial_a f_k\,\partial_b f_k$ be the (uncentered) Fisher information metric. Then
\begin{equation}
g_{ab} = \sum_{u,v\in\{0,1,2,12\}} T^{uv}_{ab}
\qquad\text{(uncentered),}
\label{eq:fs-uncentered}
\end{equation}
exactly, and the \emph{centered} Fisher information, obtained by restricting the sum to
$u,v\in\{1,2,12\}$, satisfies
\begin{equation}
\begin{aligned}
g^{\mathrm{cent}}_{ab}
&:= \sum_{u,v\in\{1,2,12\}} T^{uv}_{ab}\\
&= g_{ab}
- \Big(T^{00}_{ab}
+ \sum_{v\in\{1,2,12\}}\big(T^{0v}_{ab}+T^{v0}_{ab}\big)\Big),
\end{aligned}
\label{eq:fs-centered}
\end{equation}
i.e.\ the $u=0$ (mean) channel makes up the exact difference between the two.
\end{theorem}
\begin{proof}
The Hoeffding decomposition is a partition of unity of projection operators, $f=\sum_{u}h_u(f)$, so by
the commutation property $\partial_a f=\sum_{u}\partial_a h_u$. Substituting into
$g_{ab}=\sum_k\partial_a f_k\,\partial_b f_k$ and expanding the product of the two finite sums,
\begin{equation}
\begin{aligned}
g_{ab}
&=\sum_k\Big(\sum_u\partial_a h_u^k\Big)
        \Big(\sum_v\partial_b h_v^k\Big)\\
&=\sum_{u,v}\sum_k\partial_a h_u^k\,\partial_b h_v^k\\
&=\sum_{u,v\in\{0,1,2,12\}}T^{uv}_{ab},
\end{aligned}
\end{equation}
which is~\eqref{eq:fs-uncentered}. Isolating the terms that contain the mean index $u=0$ or $v=0$
gives $g_{ab}=g^{\mathrm{cent}}_{ab}+T^{00}_{ab}+\sum_{v\neq0}(T^{0v}_{ab}+T^{v0}_{ab})$, which
rearranges to~\eqref{eq:fs-centered}. The identity is algebraic and holds for any covariate
distribution, in particular under arbitrary covariate dependence.
\end{proof}

Define $S_u:=\operatorname{tr}(T^{uu})/\operatorname{tr}(g)$, the information-theoretic analogue of a
Sobol' index. Unlike the classical (variance-based) index, $\sum_u S_u$ is not guaranteed to equal
$1$ or to lie in $[0,1]$: the cross terms $T^{uv}$, $u\neq v$, can be of either sign, and their sign
has a precise operational meaning in terms of the fit landscape.

To make this rigorous, recall that for a weighted least-squares fit the Fisher metric $g$ is the
Gauss--Newton Hessian of the cost $C(\theta)=\tfrac12\sum_k r_k(\theta)^2$, written in terms of the
studentized residual $r_k=[Y_k-f_k(\theta)]/\sigma_k$: at a fit point,
\begin{equation}
\partial_a\partial_b C = g_{ab} - \sum_k \frac{r_k}{\sigma_k}\,\partial_a\partial_b f_k \;\approx\; g_{ab},
\label{eq:hessian}
\end{equation}
the residual term vanishing for an exact fit or a locally linear model. As in
Section~\ref{sec:identity} we work in the whitened normalization $\sigma_k\equiv1$, so that
$g_{ab}=\sum_k\partial_a f_k\,\partial_b f_k$ as in \eqref{eq:Tdef}. A parameter perturbation
$\xi\in\mathbb R^P$ therefore raises the cost by $\Delta C=\tfrac12\,\xi^\top g\,\xi=\tfrac12\|J\xi\|^2$,
where $J_{ka}=\partial_a f_k$ is the Jacobian. The Hoeffding identity $\partial_a f=\sum_u\partial_a
h_u$ splits the output perturbation into channel contributions in data space,
\begin{equation}
J\xi=\sum_u v_u(\xi), \qquad v_u(\xi)_k:=\sum_a\partial_a h_u^k\,\xi_a\in\mathbb R^N,
\label{eq:channel-output}
\end{equation}
with $\xi^\top T^{uv}\xi=\langle v_u(\xi),v_v(\xi)\rangle$ the data-space inner product of two
channels' responses. The following makes the complementary/compensating dichotomy precise.

\begin{proposition}[Sign of the cross term]
\label{prop:sign}
Fix a perturbation $\xi$ and write $s_u=\|v_u(\xi)\|^2\ge0$ and $C_{uv}=\langle v_u(\xi),v_v(\xi)
\rangle$. Then $2\,\Delta C=\sum_u s_u+2\sum_{u<v}C_{uv}$, and:
\begin{enumerate}
\item[(i)] \emph{(complementary)} if $C_{uv}>0$ the two channel responses form an acute angle and add
super-additively, $\|v_u+v_v\|^2>s_u+s_v$: the joint move is \emph{more} detectable than the sum of
the separate channels;
\item[(ii)] \emph{(compensating)} if $C_{uv}<0$ they form an obtuse angle and add sub-additively,
$\|v_u+v_v\|^2<s_u+s_v$. Concretely, given a move $p$ that excites channel $u$, the cost is lowered by
adding any move $q$ that excites channel $v$: over the family $\xi=p+tq$ the cost is minimized at
$t^\star=-\langle Jp,Jq\rangle/\|Jq\|^2$, with
\begin{equation}
2\,\Delta C(t^\star)=\|Jp\|^2\Big(1-\cos^2\phi\Big),\qquad \cos\phi=\frac{\langle Jp,Jq\rangle}{\|Jp\|\,\|Jq\|},
\label{eq:offset}
\end{equation}
so a nonzero compensating shift that strictly lowers the cost exists iff the channels overlap
($\langle Jp,Jq\rangle\neq0$), and it opposes that overlap. In the collinear limit $v_v=-v_u$ the total
response vanishes, $J\xi=0$: $\xi$ is a flat direction of the cost, an exactly unidentifiable
parameter combination.
\end{enumerate}
\end{proposition}
\begin{proof}
Expanding $\|J\xi\|^2=\|\sum_u v_u\|^2$ gives $2\Delta C=\sum_u\|v_u\|^2+2\sum_{u<v}\langle v_u,v_v
\rangle$, the stated identity.

\emph{Part (i).} The two channel responses $v_u,v_v$ are vectors in data space $\mathbb R^N$, so the
Cauchy--Schwarz inequality gives $|C_{uv}|=|\langle v_u,v_v\rangle|\le\|v_u\|\,\|v_v\|=\sqrt{s_us_v}$,
with $C_{uv}=\sqrt{s_us_v}\,\cos\phi_{uv}$ defining the angle $\phi_{uv}$ between them. Substituting
into the parallelogram law $\|v_u+v_v\|^2=s_u+s_v+2C_{uv}$ shows that the joint response is
super-additive ($>s_u+s_v$) exactly when $\cos\phi_{uv}>0$ and sub-additive when $\cos\phi_{uv}<0$;
Cauchy--Schwarz bounds the excess by $2\sqrt{s_us_v}$ in either direction. This proves (i) and the
first sentence of (ii).

\emph{Part (ii).} The compensating mechanism is the offset construction of \citet{transtrum2010,
transtrum2011}. Along the ray $\xi=p+tq$ the cost is $2\Delta C(t)=\|Jp+tJq\|^2$, a convex quadratic
in $t$ minimized at $t^\star=-\langle Jp,Jq\rangle/\|Jq\|^2$; substituting gives \eqref{eq:offset},
which lies strictly below $\|Jp\|^2$ iff $\cos\phi\neq0$. Thus whenever the two channels overlap, a
nonzero shift along $q$ strictly lowers the cost relative to moving along $p$ alone. In the collinear
limit $v_v=-v_u$ these two terms cancel in $J\xi=\sum_w v_w$; when no other channel is excited
$J\xi=0$ and $\Delta C=0$ to second order, so $\xi$ is a flat (sloppy) direction of the cost.
\end{proof}
\noindent Thus a positive cross term means the channels carry \emph{complementary} information (a
perturbation is detectable through both at once), while a negative cross term means they
\emph{compensate}: a shift routed through one channel can be partly cancelled by a shift through
another, leaving the fit almost unchanged. The extreme case is a smallest-eigenvalue eigenvector
$\hat n$ of $g$, the ``sloppy'' direction in the sense of \citet{transtrum2010,transtrum2011}, for which
$\lambda_{\min}=\sum_u\|v_u(\hat n)\|^2+2\sum_{u<v}\langle v_u(\hat n),v_v(\hat n)\rangle$ can be tiny
even when the individual channel responses $\|v_u(\hat n)\|$ are large, precisely because the cross
terms are negative and nearly cancel the diagonal sum. Negative cross terms are therefore the
channel-resolved signature of parameter sloppiness.

\subsection{Which cross terms vanish under independence}
\label{sec:vanish}
\begin{proposition}
\label{prop:vanish}
Under $X_1\perp X_2$, the population cross terms $T^{0,12},T^{1,2},T^{1,12},T^{2,12}$ vanish for every
$\theta$; together with $T^{0,1},T^{0,2}$ (which vanish by centering) this leaves only the four
diagonal channels at the population level.
\end{proposition}
\begin{proof}
Fix $X_2\perp X_1$. By definition of the interaction component, $\mathbb E[h_{12}\mid X_1]=0$
identically in $\theta$ (the Hoeffding interaction has vanishing conditional mean given either single
covariate). A function identically zero in $\theta$ has all $\theta$-derivatives zero, so
$\mathbb E[\partial_b h_{12}\mid X_1]=0$ for every $\theta,b$. Hence, using the tower property and the
$\sigma(X_1)$-measurability of $\partial_a h_1$,
\begin{equation}
T^{1,12}_{ab}=\mathbb E[\partial_a h_1\,\partial_b h_{12}]
=\mathbb E\big[\partial_a h_1\,\mathbb E[\partial_b h_{12}\mid X_1]\big]=0 .
\end{equation}
The identical argument with the roles of $X_1,X_2$ exchanged gives $T^{2,12}=0$, and applied to the
mean channel ($\partial_a h_0$ being constant) gives $T^{0,12}=0$; independence of $X_1,X_2$ gives
$T^{1,2}=\mathbb E[\partial_a h_1]\,\mathbb E[\partial_b h_2]=0$ after centering. The pairs
$T^{0,1},T^{0,2}$ vanish because $\partial_b h_{1}, \partial_b h_2$ are centered. All expectations are
population (infinite-$N$) quantities; at finite $N$ the corresponding empirical traces are nonzero
sampling fluctuations.
\end{proof}
\noindent All these are population-level detectors; without an explicit noise baseline (bootstrap), a
small nonzero cross term at finite $N$ cannot be distinguished from noise, so the framework predicts
its own finite-sample noise floor for the cross terms. Under true independence the persistent cross
terms $T^{1,12},T^{2,12}$ are therefore sampling fluctuations that decay as $N$ grows, rather than
genuine model-induced interaction (Proposition~\ref{prop:vanish}); we quantify this noise floor
directly on the datasets of Section~\ref{sec:results} through the bootstrap distributions of the
cross-block traces, against which the observed interactions are assessed.

\subsection{Estimators}
\label{sec:estimators}
The components $h_u$ can be estimated in two ways, which are asymptotically equivalent but differ in
finite-sample bias and variance. The first is a nonparametric pick-and-freeze estimator \citep{sobol1993,saltelli2010}: each
$h_u$ is obtained by re-pairing the covariates, fixing the channel of interest and averaging over the
re-paired grid of the remaining variables (context and the other covariate), with the population mean
estimated as the full $N\times N$ re-paired grid average rather than the naive paired-sample mean
(the two differ under covariate dependence, and the difference concentrates in the interaction
channel); its cost is $O(N^2)$. The second is a parametric orthogonal-basis projection \citep{sudret2008}: one
projects $\partial_a f$ onto an orthonormal basis of the $K-1$ nontrivial contrasts of each
$K$-category covariate, taken with respect to the empirical marginal measure, at cost $O(N)$. We build
this basis by a weighted-QR orthonormalization of the indicator design matrix \citep{golub2013}: with $D$ the
$K\times K$ matrix of category indicators (including the constant) and $w=\sqrt{p}$ the square-root
marginal weights, the factorization $\operatorname{diag}(w)\,D=QR$ yields, after rescaling, a set of
functions orthonormal to machine precision for any $K$; since the Fisher-Sobol channel traces are
invariant to the choice of orthonormal basis within a channel, this coding-invariant construction
gives the same indices as any other orthonormal basis, while avoiding the loss of orthonormality that
afflicts Gram--Schmidt on the categorical raw moments beyond a handful of levels. In both estimators
the context coordinate (luminosity for the luminosity functions, ambient concentration for the plant
data) is marginalized, by the re-paired grid in the nonparametric case and by the empirical average
in the parametric case, so that both yield the same four channels $\{\bar f,X_1,X_2,X_1X_2\}$; the
parametric route additionally requires the truncation order to match the true function.

\section{Selected datasets}
\label{sec:datasets}

\subsection{Plant CO$_2$-uptake experiment}
\label{sec:plant-data}
As a small, purely categorical example we use the classic plant CO$_2$-uptake dataset of
\citet{potvin1990} (distributed with \textsf{R} as \texttt{CO2}), which measures
the CO$_2$ uptake of the grass \emph{Echinochloa crus-galli} as a function of ambient CO$_2$
concentration for twelve plants at seven concentrations each. To keep the design balanced and both
covariates binary, structurally parallel to the luminosity-function example below, we restrict to
a $2\times2$ subset: the six Mississippi-origin plants, with a chilling-treatment factor
($X_1\in\{$nonchilled, chilled$\}$) and a plant-replicate factor ($X_2\in\{$replicate~1,
replicate~2$\}$) as the two categorical covariates, giving $N=28$ points across four
covariate cells. We model the uptake with a Michaelis--Menten saturation curve modulated
multiplicatively by the two factors,
\begin{equation}
f(c,X_1,X_2;\theta) = \frac{u_{\max}\,c}{K+c}\,\big(1+\theta_2 X_1+\theta_3 X_2\big),
\label{eq:plant-model}
\end{equation}
with $\theta=(K,u_{\max},\theta_2,\theta_3)$; the concentration $c$ plays the role of a fixed context
coordinate and the two binary covariates enter the ANOVA channels. Uptake values are assigned a
nominal $7\%$ relative uncertainty.

\subsection{Radio luminosity functions: the AGN--SFG model}
\label{sec:lf-data}
As a larger, astrophysically motivated case we fit a combined 1.4~GHz radio luminosity-function (LF)
model spanning two physically distinct populations, star-forming galaxies (SFG) and radio active
galactic nuclei (AGN), which are described by \emph{different} functional forms. The population type
enters as a binary covariate $X_1\in\{0,1\}$ (SFG/AGN) that switches between the two forms, and a
second binary covariate $X_2$ marks the source catalogue of origin; a genuine population$\times$%
catalogue coupling is therefore plausible on structural grounds, not merely as a statistical artifact.

The data are published binned LF determinations, $N=38$ points in total, drawn from two surveys. The
local benchmark ($X_2=0$) is the 6dFGS--NVSS local radio luminosity function of \citet{mauch2007},
measured with the $1/V_{\max}$ method \citep{schmidt1968} from a sample of 7\,824 radio sources with
$S_{1.4}\geq2.8$~mJy, $K\leq12.75$~mag and $z>0.003$ (median $\bar z\approx0.046$), of which 4\,006
star-forming galaxies and 2\,661 radio-loud AGN enter the LF; we use their tabulated values in
$0.4$~dex bins of radio power, 10 SFG bins spanning $\log L_{1.4}=20.0$--$23.6$ and 16 AGN bins
spanning $20.4$--$26.4$, converted from their $\mathrm{mag^{-1}\,Mpc^{-3}}$ normalization to
$\mathrm{dex^{-1}\,Mpc^{-3}}$ (an offset of $\log_{10}2.5$). The deep-field points ($X_2=1$) come from
the VLA-COSMOS 3~GHz Large Project, a 2~deg$^2$ survey of the COSMOS field with
rms~$\approx2.3\,\mu$Jy\,beam$^{-1}$: the SFG luminosity function of \citet{novak2017} (based on
$\sim$6\,000 star-forming galaxies with optical--NIR counterparts) and the radio-AGN luminosity
function of \citet{smolcic2017} (based on $\sim$1\,800 radio-excess AGN), from which we take the
lowest redshift bin, $0.1<z<0.4$ (median $z\approx0.31$), with 6 points per population; luminosities
are rest-frame 1.4~GHz values obtained in those works from the observed 3~GHz fluxes with measured
spectral indices. All quoted LF uncertainties are Poisson-based and asymmetric (small-number
confidence intervals following \citealt{gehrels1986} in the sparsely populated bins); where needed we
symmetrize them as the mean of the upper and lower errors. Because the local and COSMOS
determinations refer to slightly different epochs ($z\approx0$ vs.\ $z_{\mathrm{med}}\approx0.31$),
the catalogue covariate $X_2$ absorbs both survey-to-survey systematics and this modest redshift
offset; we make no attempt to model LF evolution explicitly.

The published uncertainties are strongly heteroscedastic, ranging from $0.01$~dex in the
best-populated local bins to $0.56$~dex at the bright end, a dynamic range of $\sim3\times10^3$ in
the weights $w_k=1/\sigma_k^2$. The model parameters are therefore estimated by \emph{weighted}
(inverse-variance) least squares using the symmetrized uncertainties; the resulting fit quality,
$\chi^2/\mathrm{dof}\approx3.5$, indicates that the 10-parameter form is an approximate (though,
as we show below, informative) description of the heterogeneous published LFs. For the
\emph{decomposition} itself we use the unweighted metric. The fully weighted Fisher metric is
available exactly within the framework (the weighted Gram matrix of Sect.~\ref{sec:identity}), but is
not informative here: inverse-variance weighting concentrates essentially the entire metric on the few
most precise local mid-luminosity bins, reducing the channel decomposition to a near-cancellation
between the mean and $X_1$ channels rather than a description of the full data set. With the
unweighted metric every luminosity bin contributes on an equal footing, which matches the intent of
the illustration.

The radio LF of star-forming galaxies is not well described by a pure Schechter function \citep{schechter1976}; instead
the standard form is the ``power-law plus lognormal'' of \citet{saunders1990},
\begin{equation}
\begin{aligned}
S(L;\Phi_0,L_\star,\alpha,\sigma)
&= \Phi_0\Big(\frac{L}{L_\star}\Big)^{1-\alpha}\\
&\quad\times
\exp\!\Big[-\frac{1}{2\sigma^2}\log_{10}^2\!\Big(1+\frac{L}{L_\star}\Big)\Big],
\end{aligned}
\label{eq:schechter}
\end{equation}
where $L_\star$ sets the turnover luminosity, $\Phi_0$ the normalization, and $\alpha,\sigma$ the
faint- and bright-end shapes. The local 1.4~GHz SFG LF was measured by \citet{mauch2007}
(6dFGS--NVSS) with $\Phi_0=10^{-2.83}\,\mathrm{mag^{-1}Mpc^{-3}}$, $L_\star=10^{21.18}\,
\mathrm{W\,Hz^{-1}}$, $\alpha=1.02$, $\sigma=0.60$, values we adopt as the SFG-branch initialization;
the same form was fit to the deep VLA-COSMOS 3~GHz sample and its cosmic evolution by
\citet{novak2017}. The radio-AGN LF instead follows a broken (double) power law, analogous to the optical quasar LF
\citep{dunlop1990,willott2001,brown2001},
\begin{equation}
D(L;\Phi_0,L_\star,\alpha,\beta) = \frac{\Phi_0}{(L_\star/L)^{\alpha}+(L_\star/L)^{\beta}},
\label{eq:doublepl}
\end{equation}
with faint- and bright-end slopes $\alpha,\beta$. \citet{mauch2007} fit the local radio-AGN LF with
$\Phi_0=10^{-5.50}\,\mathrm{mag^{-1}Mpc^{-3}}$, $L_\star=10^{24.59}\,\mathrm{W\,Hz^{-1}}$,
$\alpha=1.27$, $\beta=0.49$, which we adopt as the AGN-branch initialization; the cosmic evolution of
the radio-AGN LF to $z\sim5$ was characterized on the VLA-COSMOS 3~GHz sample by
\citet{smolcic2017}.

Writing $L=10^{x}$, we couple the catalogue covariate $X_2$ \emph{multiplicatively} inside the
logarithm,
\begin{equation}
\begin{aligned}
f(L,X_1,X_2;\theta) = \log_{10}\Big\{&\big[S(L;\theta_{0:4})(1-X_1)+D(L;\theta_{4:8})X_1\big]\\
&\times\big(1 + \theta_8 X_2(1-X_1) + \theta_9 X_2 X_1\big)\Big\},
\end{aligned}
\label{eq:lf-model}
\end{equation}
a 10-parameter nonlinear model, with all partial derivatives obtained analytically (verified against
finite differences to $\sim10^{-9}$ relative error). Multiplicative rather than additive coupling is
essential here: an additive covariate term inside the logarithm would make the interaction derivative
scale as the inverse of the bracketed base term, which diverges at the bright end where the LF
vanishes and concentrates all leverage on a few points, whereas the multiplicative form keeps the
covariate derivatives $O(1)$ everywhere and confines the bright-end leverage to the base-parameter
channels. The fit is constrained
to the positive-modulation region $1+\theta_8 X_2(1-X_1)+\theta_9 X_2 X_1>0$, which holds comfortably
at the fitted values.

\section{Results}
\label{sec:results}
For each dataset we show three diagnostics computed by the identical pipeline: the per-block
eigenvalue spectrum of $T^{uv}_{ab}$, the Fisher-Sobol contributions $S_u$ (and the cross term), and
the bootstrap distribution of each channel's contribution under repeated train/test splitting,
comparing the nonparametric (pick-and-freeze) and parametric (orthogonal-basis) estimators of
Section~\ref{sec:estimators}.

\subsection{Plant CO$_2$ uptake}
\label{sec:plant-results}
\begin{figure*}[tp]
\centering\includegraphics[width=\textwidth]{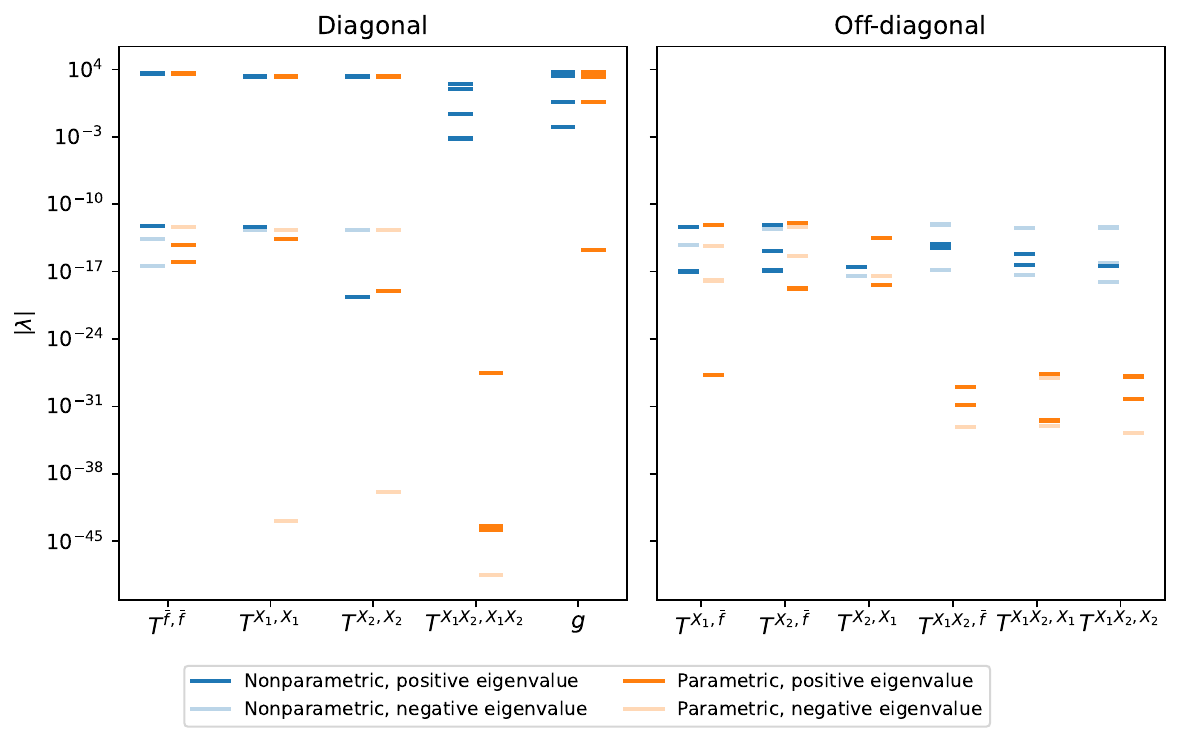}
\caption{Plant CO$_2$-uptake fit: eigenvalue spectrum of $T^{uv}_{ab}$, diagonal blocks (left) and
off-diagonal blocks (right). Blue/orange $=$ nonparametric/parametric estimator; light shades mark
negative eigenvalues.}
\label{fig:plants-spectrum}
\end{figure*}

\begin{figure*}[tp]
\centering\includegraphics[width=0.42\textwidth]{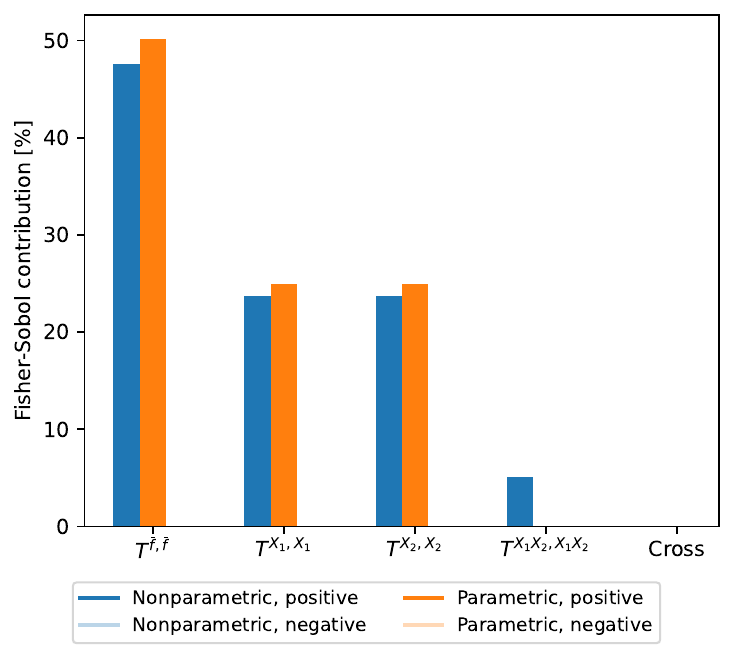}
\caption{Plant CO$_2$-uptake fit: Fisher-Sobol contributions, nonparametric (blue) vs.\ parametric
(orange). Information is carried by the mean level and, roughly equally, the treatment covariate $X_1$
and the replicate covariate $X_2$, with a small interaction; the two estimators agree.}
\label{fig:plants-bar}
\end{figure*}

\begin{figure*}[tp]
\centering\includegraphics[width=\textwidth]{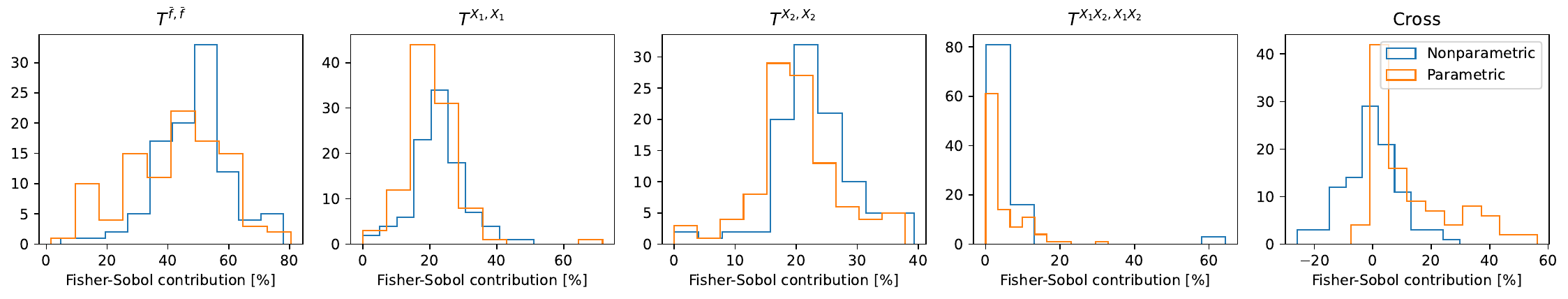}
\caption{Plant CO$_2$-uptake fit: bootstrap distributions of the per-channel Fisher-Sobol
contributions ($100$ resamples). The diagonal channels agree; the parametric interaction/cross
channels are modestly wider and slightly positive, consistent with the finite-sample noise floor of
Section~\ref{sec:vanish}.}
\label{fig:plants-bootstrap}
\end{figure*}

Figures~\ref{fig:plants-spectrum}--\ref{fig:plants-bootstrap} show the three diagnostics for the plant
model~\eqref{eq:plant-model}. On the full sample the Fisher information is dominated by the mean level
($\bar f\approx48\%$) and split almost equally between the treatment covariate $X_1$ and the
replicate covariate $X_2$ ($\approx24\%$ each), with a small interaction ($\lesssim5\%$) and a
negligible cross term (Figure~\ref{fig:plants-bar}); the two estimators agree to within $\sim5$
percentage points on every channel. Under bootstrap resampling
(Figure~\ref{fig:plants-bootstrap}) the diagonal channels remain in close agreement, while the
parametric estimator shows a modestly wider and slightly positive interaction/cross channel than the
nonparametric one. Both covariates are binary here, so no leverage pathology of the kind seen for the
log-space luminosity-function model can arise; the residual spread is the finite-sample noise floor of
Section~\ref{sec:vanish}, on the parametric side, and it shrinks toward zero as the evaluation sample
grows. This is the mirror image of the luminosity-function example: there the nonparametric estimator
was the more fragile one (bright-end leverage), whereas here the parametric orthogonal-basis
projection is, since it estimates the interaction contrast from the few points falling in each
covariate cell; neither estimator dominates, and their agreement is the diagnostic that the
interaction channel is genuinely small.


\subsection{Radio luminosity functions}
\label{sec:lf-results}
\begin{figure*}[tp]
\centering\includegraphics[width=\textwidth]{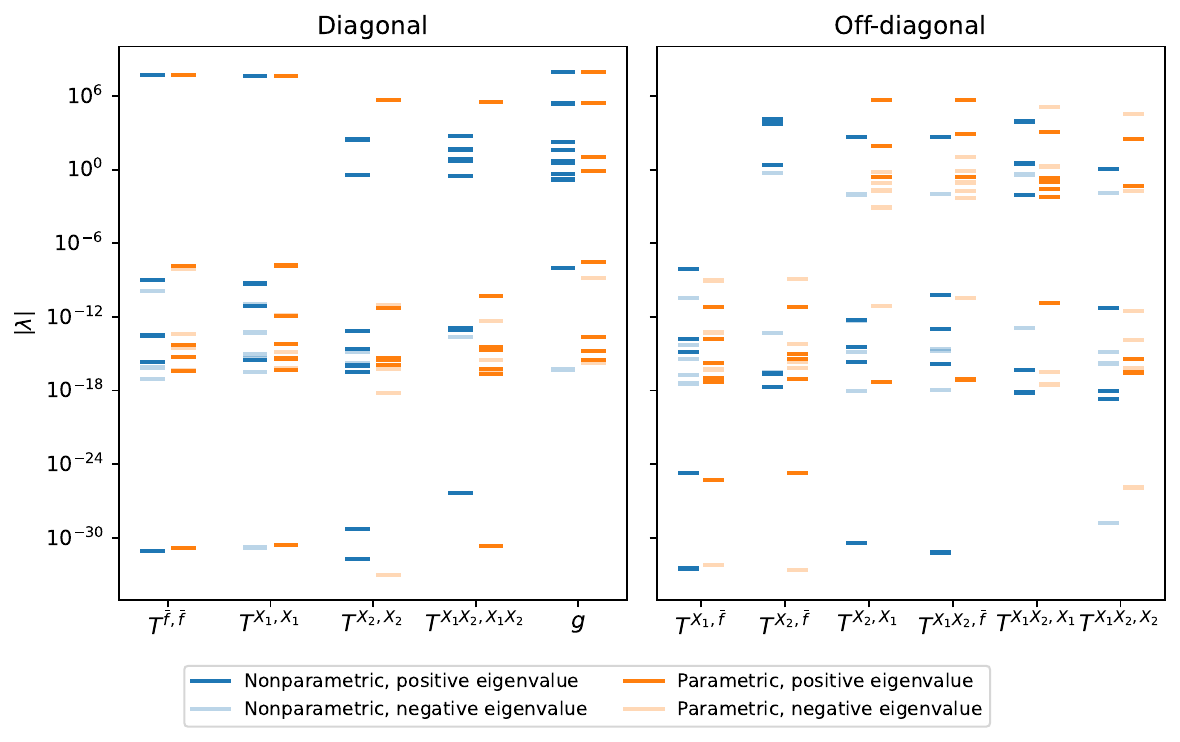}
\caption{Radio luminosity-function fit: eigenvalue spectrum of $T^{uv}_{ab}$, diagonal (left) and
off-diagonal (right) blocks. The spectrum is dominated by the $\bar f$ and $X_1$ blocks.}
\label{fig:lf-spectrum}
\end{figure*}

\begin{figure*}[tp]
\centering\includegraphics[width=0.42\textwidth]{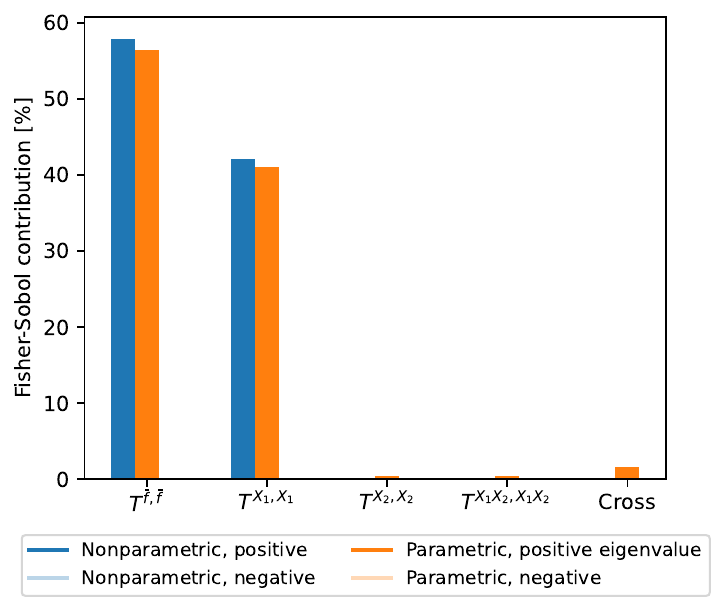}
\caption{Fisher-Sobol contributions for the LF fit, nonparametric (blue) vs.\ parametric (orange).
The information is carried by the mean level and the functional-form covariate $X_1$; the catalogue
covariate $X_2$ and the interaction are small.}
\label{fig:lf-bar}
\end{figure*}

\begin{figure*}[tp]
\centering\includegraphics[width=\textwidth]{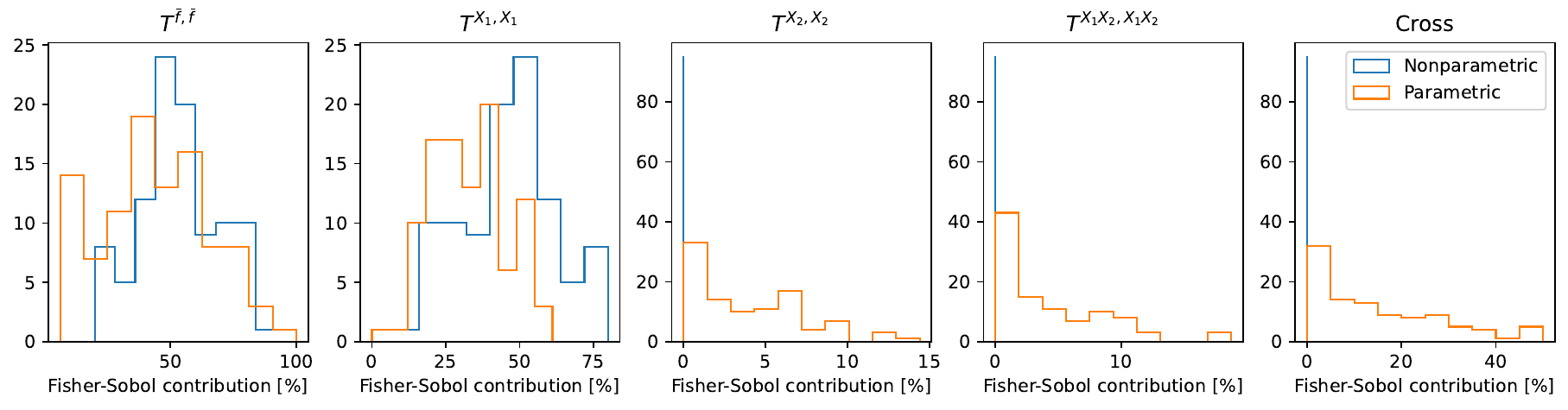}
\caption{Bootstrap distributions of the per-channel Fisher-Sobol contributions for the LF fit
($100$ resamples). Both estimators are stable and bounded; the interaction and cross channels
concentrate near zero, with no heavy tails.}
\label{fig:lf-bootstrap}
\end{figure*}

Figures~\ref{fig:lf-spectrum}--\ref{fig:lf-bootstrap} show the same three diagnostics for the AGN--SFG
model~\eqref{eq:lf-model}. Three features stand out. First, the Fisher information is overwhelmingly
carried by the mean level and the functional-form covariate $X_1$: on the full sample the
nonparametric estimator assigns $(\bar f,X_1,X_2,X_1X_2,\text{Cross})\approx(58,42,0,0,0)\%$ and the
parametric estimator $\approx(56,41,0.5,0.4,1.6)\%$ (Figure~\ref{fig:lf-bar}). Second, the two
estimators agree: the wild estimator-to-estimator discrepancy produced by additive coupling is
absent. Third, the catalogue covariate $X_2$ and the interaction carry a small but genuinely nonzero
share (the multiplicative modulation is real: the weighted fit gives $\theta_8\approx0.87$,
$\theta_9\approx0.29$, and dropping the $X_2$ terms raises the weighted $\chi^2$ from $99$ to $254$,
i.e.\ $\Delta\chi^2\approx155$ for two parameters); the parametric estimator,
which projects onto an orthonormal covariate basis, surfaces this share, while the nonparametric
estimator (measuring absolute Fisher information, which is dominated by the parametrization scale of
the shape parameters) rounds it to zero. This residual difference is a parametrization-scale effect,
not the qualitative disagreement of the additive model. Under bootstrap resampling
(Figure~\ref{fig:lf-bootstrap}) both estimators are stable and bounded, with no heavy tails.

\begin{figure*}[tp]
\centering\includegraphics[width=\textwidth]{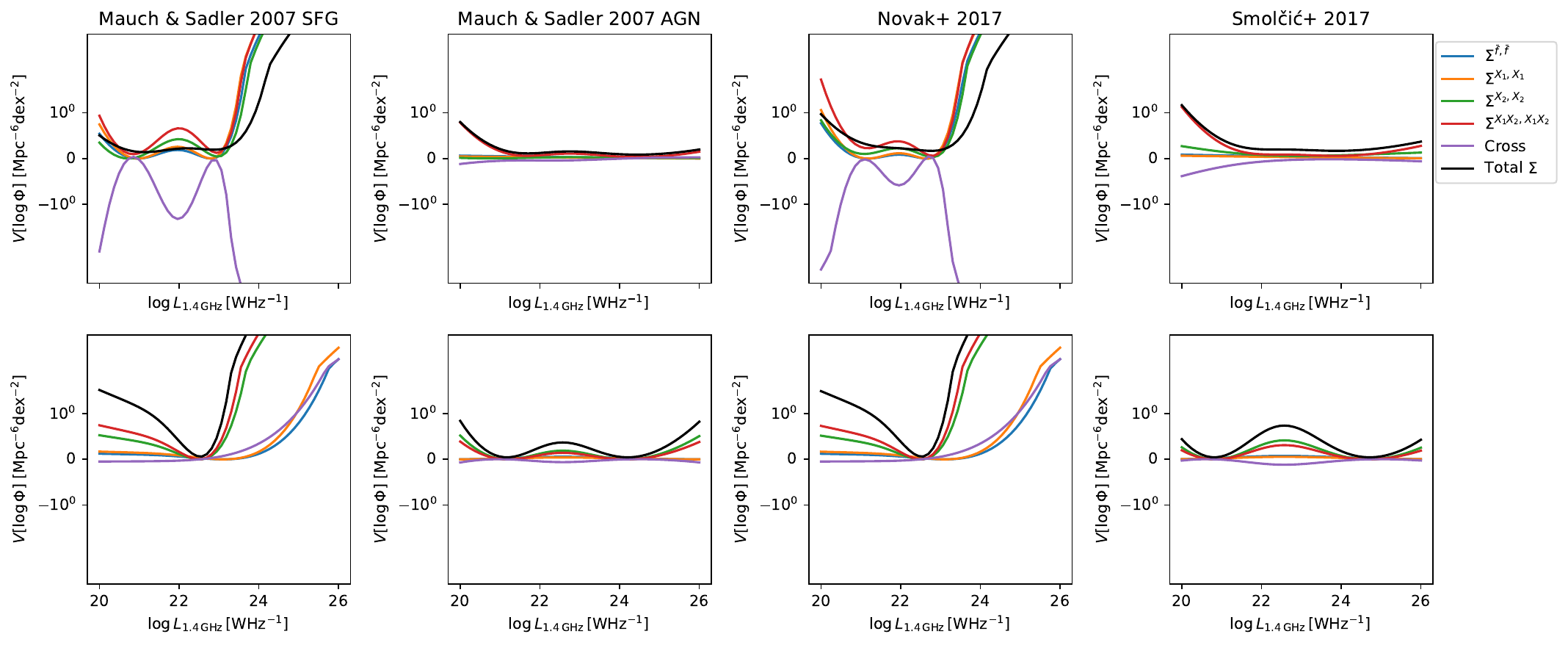}
\caption{Pulled-back Fisher-Sobol decomposition of the radio luminosity-function model: the
contribution $\Sigma^{uv}(L)=J g^{+}T^{uv}g^{+}J^\top$ of each channel block to the Cram\'er--Rao
variance of the model prediction (the fitted $\log\Phi$) as a function of luminosity $L$, for the
four population$\times$catalogue corners (columns) and for the nonparametric (top) and parametric
(bottom) estimators. Curves: the diagonal channels $\bar f$, $X_1$, $X_2$, $X_1X_2$; the signed sum
of the off-diagonal terms (Cross); and the total $\Sigma_{\mathrm{tot}}=J g^{+}J^\top$ (black), to
which the channel curves sum exactly. The two star-forming (SFG) catalogues are the local 6dFGS--NVSS fit of
\citet{mauch2007} and the VLA-COSMOS 3~GHz fit of \citet{novak2017}; the two AGN catalogues are the local fit of
\citet{mauch2007} and the VLA-COSMOS 3~GHz fit of \citet{smolcic2017}.}
\label{fig:lf-pulledback}
\end{figure*}

A complementary, spatially resolved view is given in Figure~\ref{fig:lf-pulledback}, which pulls the
channel decomposition back onto the observable. With $J_a(L)=\partial f/\partial\theta_a$ evaluated
along a luminosity grid at fixed $(X_1,X_2)$, we plot the contribution of each channel block to the
Cram\'er--Rao variance of the predicted $\log\Phi$,
\begin{equation}
\Sigma^{uv}(L) \;=\; J_a(L)\,\big[g^{+}\,T^{uv}\,g^{+}\big]^{ab}\,J_b(L),
\label{eq:pullback}
\end{equation}
where $g^{+}$ denotes the Moore--Penrose pseudo-inverse of the full metric $g=\sum_{uv}T^{uv}$.
Because $g^{+}g\,g^{+}=g^{+}$, these contributions sum \emph{exactly} to the total Cram\'er--Rao
variance of the prediction, $\sum_{uv}\Sigma^{uv}(L)=J\,g^{+}J^\top=:\Sigma_{\mathrm{tot}}(L)$:
Eq.~\eqref{eq:pullback} distributes the prediction variance over the Fisher-Sobol channels with no
remainder, the spatially resolved counterpart of Theorem~\ref{thm:fs}. The diagonal contributions
$\Sigma^{uu}$ are nonnegative (each $g^{+}T^{uu}g^{+}$ is positive semi-definite); the ``Cross''
curve, the same construction applied to the summed off-diagonal blocks $\sum_{u\neq v}T^{uv}$, is
indefinite and carries the sign of the channel interplay. Each panel of
Figure~\ref{fig:lf-pulledback} shows these curves for one of the four population$\times$catalogue
corners (columns), for the nonparametric (top row) and parametric (bottom row) estimators. Two
things are visible that the integrated indices cannot show. First, the fidelity of the fitted
luminosity function is strongly luminosity-dependent: the total Cram\'er--Rao variance (black) is
smallest near each population's turnover, where the data are densest and most precise, and grows
steeply toward the faint and especially the bright end, where the parameter sensitivities $|J|$ are
largest: the bright-end extrapolation is the least constrained part of the fitted LF. Second, in the
star-forming panels of the nonparametric decomposition the cross term (purple) is \emph{negative}
over most of the luminosity range, with the diagonal channel curves individually exceeding the total
they must sum to; the negative Cross absorbs the excess, directly visualizing the compensating
(sloppy) channel combinations of Section~\ref{sec:theorem}: a shift routed through one channel is
partly cancelled by another with little change in the predicted LF, so the channels overshoot
individually and cancel in the sum. In the parametric decomposition, whose orthonormal-basis
channels are closer to mutually orthogonal in data space, the Cross stays near zero through the
well-sampled luminosities and turns positive (complementary) only in the bright-end extrapolation
region.

The astrophysical reading is concrete. For the star-forming population the bright-end variance budget
is dominated by the $X_2$ and interaction channels: what limits the bright end of the SFG luminosity
function in this combined fit is the cross-calibration between the local 6dFGS--NVSS and the deep
VLA-COSMOS determinations, not the Saunders-form shape parameters themselves. For the AGN double
power law, which has no sharp knee inside the sampled luminosity range, the information is spread far
more evenly in luminosity and the survey-coupling channels remain subdominant, with only a mild
faint-end contribution from the interaction channel. The compensating (negative-Cross) behaviour is
correspondingly a property of the star-forming branch: it straddles the turnover, where
$L_\star$--$\Phi_0$-type compensations operate, and deepens steeply into the bright-end
extrapolation, while for the AGN branch, with no sharp knee to anchor such degeneracies, the Cross
remains close to zero throughout.


\FloatBarrier
\subsection{Discussion and limitations}
\label{sec:discussion}
The decomposition in Theorem~\ref{thm:fs} is exact and model-agnostic, but the ANOVA components
themselves are not unique: they depend on the choice of channel structure, estimator, and (for the
parametric route) basis and truncation order. The method does not establish causality, and a nonzero
interaction or cross term is a statement about the fitted model's parameter information, not
necessarily about the data-generating process. A useful feature of the framework is that it predicts
its own finite-sample noise floor for the cross terms (Section~\ref{sec:vanish}), which gives a
principled way to judge whether a given interaction signal is worth interpreting. The
luminosity-function example additionally illustrates a modelling lesson specific to log-space fits:
how a covariate is coupled inside the logarithm (additively vs.\ multiplicatively) determines whether
the interaction channel is well conditioned or dominated by a few high-leverage points, the
multiplicative coupling being the better-conditioned choice.

\FloatBarrier
\section{Conclusion}
\label{sec:conclusion}
We have shown that the Fisher information metric of a fitted nonlinear model decomposes exactly across
the Hoeffding--Sobol' ANOVA channels of its covariates (Theorem~\ref{thm:fs}), giving an
information-theoretic analogue of Sobol' indices together with signed cross-channel terms whose
population-level vanishing under covariate independence (Proposition~\ref{prop:vanish}) supplies a
built-in finite-sample noise floor. Two estimators, nonparametric pick-and-freeze and parametric
orthogonal-basis projection, realize the decomposition and agree on the two categorical-covariate
datasets studied, provided log-space models couple their covariates multiplicatively so that the
interaction channel stays well conditioned. The construction turns the Fisher information matrix from
a single opaque object into an interpretable, channel-resolved accounting of where a model's parameter
information comes from.

\end{document}